\documentclass{article}

\usepackage{comment} 
\usepackage{amsmath,mathpazo}  

\usepackage{float}
\usepackage{graphicx}

\usepackage{arxiv}
\usepackage[numbers]{natbib}
\usepackage[utf8]{inputenc} 
\usepackage[T1]{fontenc}    
\usepackage{hyperref}       
\usepackage{url}            
\usepackage{booktabs}       
\usepackage{amsfonts}       
\usepackage{nicefrac}       
\usepackage{microtype}      
\usepackage{lipsum}		
\usepackage{graphicx}
\usepackage{doi}

\usepackage{mdframed}

\title{Beating Betz's Law: A larger fundamental upper bound for wind energy harvesting }

\date{September 9, 2021}	

\author{ \href{https://orcid.org/0000-0003-3639-4673}{\includegraphics[scale=0.06]{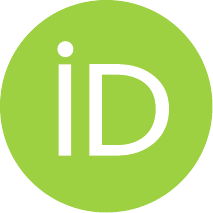}\hspace{1mm}Charlie E. M. Strauss} \\
	https://orcid.org/0000-0003-3639-4673\\	
	Browndogs Polytechnic Research Institute\\
	Los Alamos NM, 87544 \\
	\texttt{cems@browndogs.org} \\
}

\renewcommand{\shorttitle}{   C. E. M. Strauss: Beating Betz's Law}

\hypersetup{
pdftitle={BeatingBetzLaw},
pdfsubject={physics.AS},
pdfauthor={Charlie E. M. Strauss},
pdfkeywords={Wind, Turbine, Energy, Betz},
}

\begin{document}
\maketitle

\begin{abstract}
	Betz's law, purportedly, says an ideal wind harvester cannot extract more than 16/27 ($\sim$59\%)  of the wind energy.  As the law's derivation relies on momentum and energy conservation  with incompressible flow and not the physical mechanism coupling the wind-field to the extraction of work  it is ubiquitously regarded as a "universal" upper bound on efficiency, as inclusion of mechanics, aerodynamics and thermodynamics are presumed to worsen this upper bound.   Here we show that when unneeded assumptions in the Betz's law derivation are relaxed a higher bound of 2/3 ($\sim$67\%) can be achieved. A concrete example, strictly obeying the identical  energy and momentum conservation used to derive the Betz's law, is given that violates Betz's law by achieving our higher 2/3 bound.   Thus Betz's law is not a universal limit on wind energy harvesting efficiency. More surprisingly, we show Betz law is not simply the limit case of a vanishingly-thin turbine either. In 2-D models specific for turbines, radial flow is known to occur to occur as a consequence of angular momentum,\cite{Sharpe2004} but here we show in a 1-D modelthat allowing any radial flux out of the harvester cross-section can increase the efficiency without any need to consider angular momentum or explicit 2-D models. A key design insight we glean is that for high-efficiency harvesters it is better to strive for the least pressure build up (to increase flux) -- the exact opposite of the Betz model's sole operational principle of high pressure differentials.  Additionally, we derive an alternative metric of harvester efficiency which takes into account the downstream wake expansion ignored by the conventional definition of power conversion factors, and the resulting upper bound this places on power extraction from dense grids of harvesters.
\end{abstract}


\section{Prologue}

This paper is obviously counter to the common wisdom and thus a skeptic may presume it must have errors in math or physics. While that could still be a possibility, we think it is error-free as far as any 1-D derivation can be faithful to the incompressible fluid physics. Instead, we identified a hidden assumption in the canonical problem statement for Betz’s law that, when removed, surprisingly still permits an analytic solution; it also gave a new, higher, upper bound on efficiency. Moreover, the assumption we removed is one that is not physically enforced in most wind machines, so its removal is warranted. 

To persuade ourselves, let alone the reader, that such a well entrenched “law” was now broken, we constructed a concrete relevant example of a 1-D wind machine that would obey the new law and exceed Betz's law. In fact the new derivation is shorter and easier to follow and so most of the text is devoted to alternative ways of validating the new law is the correct one and Betz's law is not. In addition to proof-by-construction we also break down the mathematics step-by-step to identify the point of divergence and explain its specific physical interpretation as an unneeded, and thus mistaken, assumption that was hiding in the reduction of the wind harvester to an “actuator plane” (as done in Betz's law derivations). Like Betz and Lanchester, nowhere do we assume specific physical characteristics for the horizontal axis wind machine, so the result is even more universal than the original law.

We note however, it’s unlikely numerical simulation or physical embodiment can come close to either this upper bound or Betz’s bound. While both this derivation and Betz’s are 1-D, they implicitly include expansion of the axial wind column transverse to this axis. In any numerical simulation this requires 2-D (or 3-D) modeling because expansion must have non-uniform radial velocities (to conserve flow) and also radial kinetic energy to move the flow or work against a pressure field.\citep{conway} Those ideal 2-D physics reduce efficiency further, even ignoring any losses due to non-ideal fluids (friction, turbulence, or actuator imperfections). Thus while neither this nor Betz's law apply to a 3-D fluid, it is never-the-less true that the 1-D upper bound is still an upper bound on 3-D. The fascination of the 1-D upper bound is simply because, as far as we are aware, there is no analytic derivation of a universal upper bound in higher dimensions. Indeed, most textbooks the transition from 1-D to 2-D models only begins with considerations specific to turbines, like angular momentum or blade-tip speeds. 

Thus the bottom line is this: the following shows that if one adheres strictly to the 1-D paradigm used in Betz's law, and one deletes an unneeded assumption then one obtains a different functional form law with a higher upper bound on efficiency and different optimal operating point. But there’s greater significance here than simply out-doing Betz’s bound: 
 1. The new law’s recommendations for improving windmill design are quite different –nearly opposite– to what Betz's law recommends.
 2. It removes the absurd physical impossibilities that occur in Betz’s model when the wind machine is operating far from its optimal operating point and gives different optimal operating points when packing windmills in adjacent or stacked configurations.

\section{Introduction}
\begin{figure}[hp]
\centering
\includegraphics[scale=0.2]{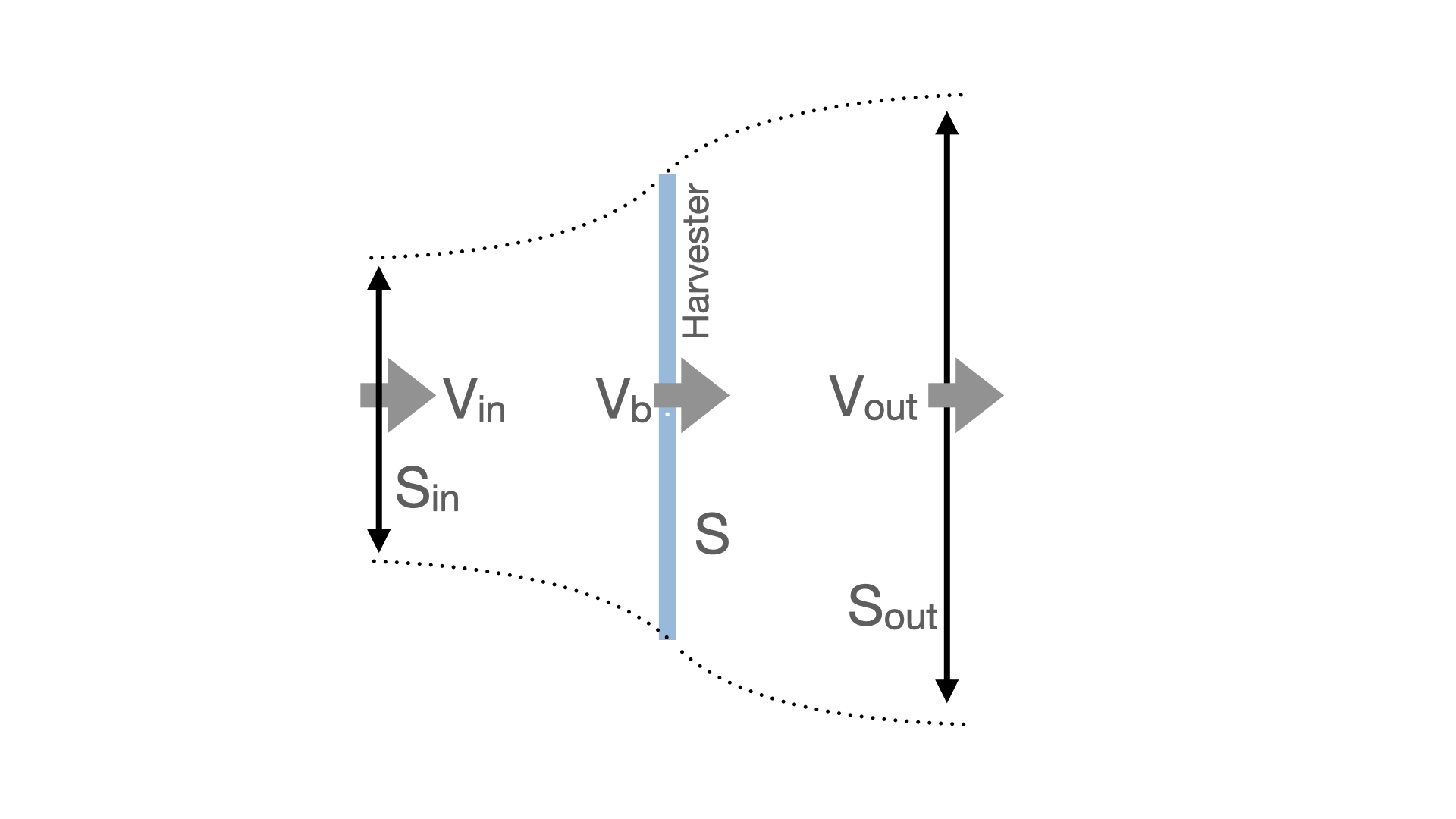}
\caption{The Canonical Betz diagram, similar to those found in the literature.  The outer curved lines are boundaries of a conserved mass flow before during and after: $\hat{m} = \; v_{in}S_{in} \; = \;v_bS \;= \;v_{out}S_{out}$. The harvester element can be anything,  but the derivation assumes the axial airflow velocity ($v_{b}$) is constant in the harvester's constant cross-sectional area ($S$), and  uniform transversely as well. These constant values place physical restrictions on how a harvester.  } \label{fig:betzmodel1}
\includegraphics[scale=0.25]{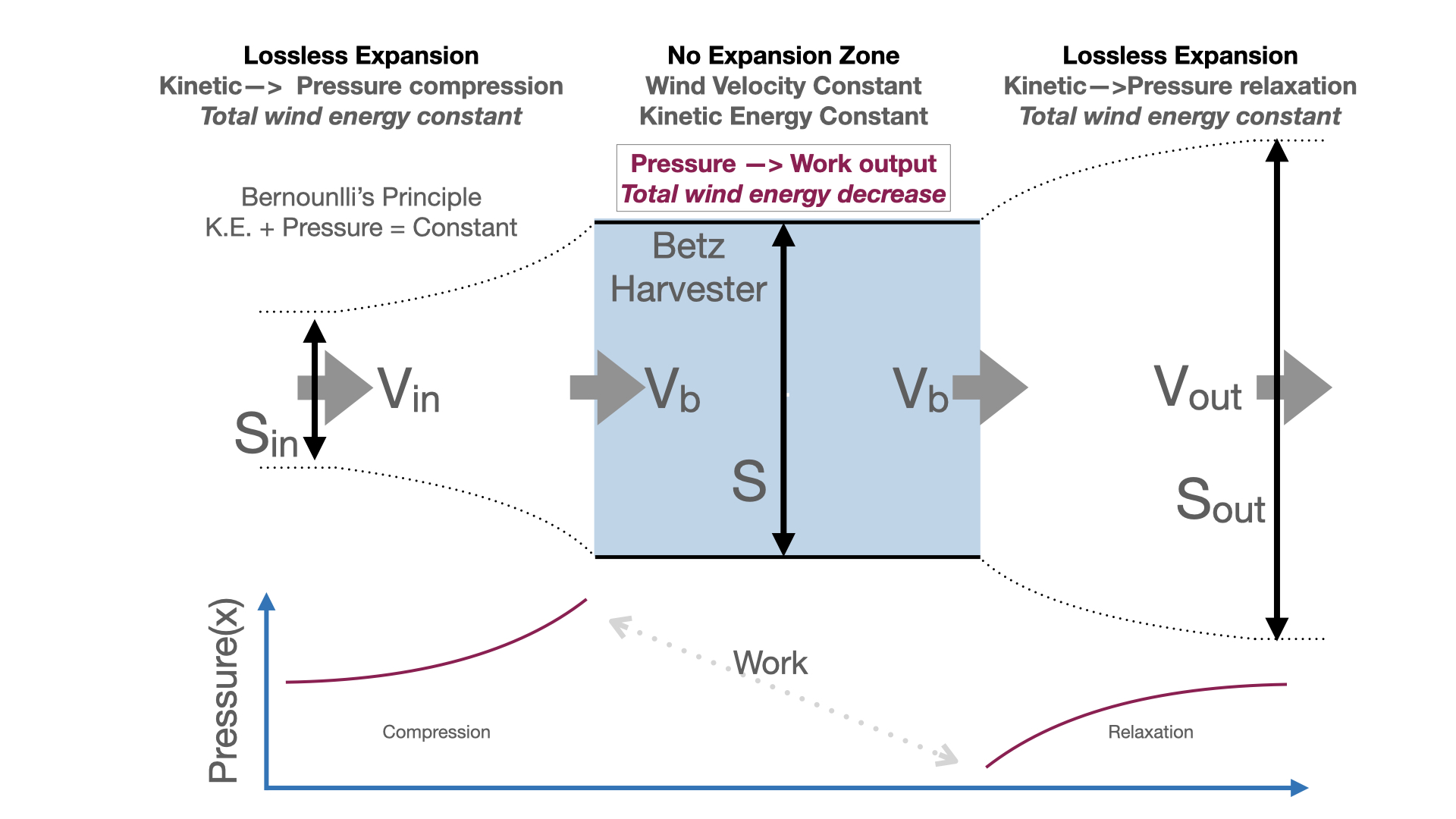}
\caption{An enhanced diagram of the Betz-type model showing the how the energy storage is shifted and extracted.  The outer curved lines represent a conserved mass flow ($\hat{m}$) before during and after: $\hat{m} = \; v_{in}S_{in} \; = \;v_bS \;= \;v_{out}S_{out}$. The harvester element can be anything, but the derivation assumes the axial airflow velocity ($v_{b}$) is constant in the harvester's constant cross-sectional area ($S$) (and  uniform transversely).  This separation of iso-energetic expansion and pressure-to-work conversion, places significant physical restrictions on how a Betz harvester can operate. Thus the Betz-type model is not universal to all possible wind harvesters, and a different upper bound is possible. We show in the main text that the Betz law is also not the limit of shrinking this back to a thin disc, and that there must be an (effective) cowling on the shaded harvester element in any Betz-compliant harvester. }\label{fig:thickbetz}

\end{figure}

When harvesting energy from flowing incompressible fluid  the Lanchester-Betz law holds that an ideal  harvester can extract no more than 16/27 ($\approx 59\%$) of the kinetic energy in a flow of the same cross-sectional area as the harvester aperture.\citep{glauert,Wind_energy_handbook,bianchi_wind_2007}  The derivation of the law idealizes the windmill to an "actuator disc"\citep{RANKINE_1865} and makes no apparent assumptions about its actual mechanism.\citep{Wind_energy_handbook,bianchi_wind_2007} (See Figure \ref{fig:betzmodel1}) Thus a rotor with infinite blades, or micro flaps, or something with no moving parts such as an electrostatic repeller in ionic wind are all conceptually upper bounded in power extraction efficiency.  The derivation relies on just conservation of energy and  1-D momentum (mass flux) under conditions of incompressible flow.  Based on just fundamentals, it is therefore ubiquitously\citep{bianchi_wind_2007, ragheb_wind_2011,renewableUK,DutchWind} regarded as a universal upper bound on windmill efficiency. Its supposed fundamental universality is held in such high esteem that some refer to is as the "carnot cycle" of horizontal axis wind machines.\citep{betzjoukowsky,ragheb_wind_2011}  Accordingly, it has become the \textit{de facto} comparison point for real world horizontal axis windmill (HAWT) performance benchmarks.\citep{ubiquity,sorensen_2015}  It stands uncontested because no real world HAWT windmill has exceeded the 59\% upper with the current best near 50\% under proper  circumstances and many, currently deployed, in the 30\% to 40\% range.\citep{Wind_energy_handbook,sorensen_2015}

Unfortunately, a \textit{gedanken} experiment shows it is not universal, and worse may not even be applicable to common wind machine mechanisms. Since the ideal Betz-type wind machine leaves over 40\% of the kinetic energy remaining in a uniform wind-field, a second identical diameter wind machine placed serially downwind after the first will extract additional power.  Since 1) every molecule of air the second machine processes was in the original input, 2) there is no force coupling between the machines in the airflow, and 3) there is no other wind source adding fresh kinetic energy between the two stages, we can view this tandem machine as one single wind harvester, as shown in Figures  \ref{fig:morph} \& \ref{fig:betzstages}. Since Betz's upper bound is violated for this construct, the law would contradict itself if the presumed universality were true. 

Because Rankine's 1865 "actuator disc," shown in Figure \ref{fig:betzmodel1}, is notionally an infinitely thin energy harvester, one might suppose that the Betz bound is simply an asymptotic limit  of a "thin" harvester.\citep{RANKINE_1865} The actuator disc model was used by Froude  in 1889, Lanchester in 1915, Betz in 1920, Joukowsky in 1920, Hoff in 1921,  and others to develop the upper bound now known as "Betz's law" (or "Joukowsky's law" or "Lanchester's law")\citep{betzjoukowsky,FROUDE,lanchester,Betz_1920,Joukowsky_1920,Hoff_1921}.  Some these original derivations, don't always mention "universal"  limits \textit{per se}, but do claim the bound is valid for thin harvesters.  
 
We claim that is also not correct.  We will show Betz's law is not the limit case of a thin wind harvester. Instead Betz law is instead a direct consequence of assumption that all the wind entering the front of the harvester exits at the back of the harvester and none escapes to the sides. That is, it behaves as though there is a cowling.  

\begin{figure}[htbp]
\centering
\includegraphics[scale=0.25]{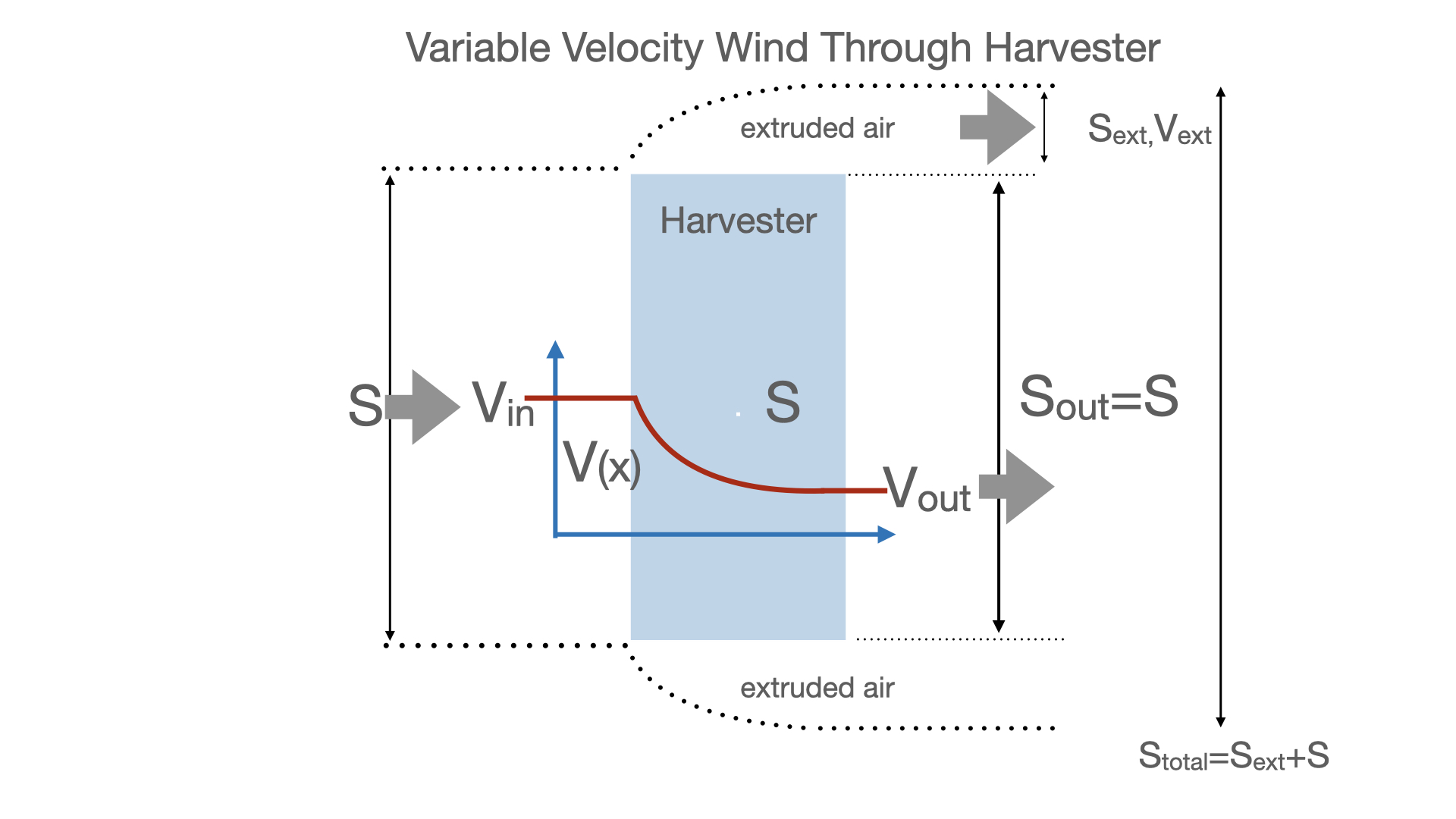}
\caption{An abstract diagram where the velocity varies continuously within the harvester. The outer lines are not cowlings but simply denote the conserved expanding airflow.  Comparing this to the Betz-type model (Fig.\ref{fig:thickbetz}), the key difference is our model incorporates the formerly external wind expansion zones in the harvester region, thus work extraction and expansion can occur at the same time.   Inside the harvester (the blue work extraction element) the  cross-section remains $S$ throughout but the wind speed along the axis is allowed to vary.  An inset graph figuratively shows a notional diminishing trajectory $v(x)$ for the velocity along the wind axis. To conserve the mass flux the envelope grows as the average speed drops across the harvester.  Thus some airflow is extruded outside the harvester and its wind energy is no longer accessible by the work extraction element. Note that the mass flux within $S$ is not conserved since it is being extruded, however the total mass flux (internal and external) remains duly conserved. While the illustration also labels the extruded air volume with an area ($S_{ext}$) and axial speed ($v_{ext}$), these are merely convenience labels on the illustration: the model itself places no restrictions on their values, location, or uniformity-- the model simply requires the extruded air conserve the mass flux. As this remains a 1-D model, no radial or tangential velocity is implied by the extrusion. }\label{fig:cemsmodel} 
\end{figure}

 We derive a new model (Fig. \ref{fig:cemsmodel}) that removes this "cowling" and we find a different maximum power extraction curve that is everywhere higher than Betz's law.  
  \begin{itemize}
\item The CEMS applies to any thickness harvester, including an "actuator disc," so Betz's law is neither universal nor even the limit of a thin actuator.
  \begin{itemize}

  \item  But with a cowling restriction, the CEMS reverts to the Betz conversion factor. This, not thinness, is the hidden assumption in the Betz law. 
     \end{itemize}
   \item The maximum conversion factor  is 2/3 (~67\%) and the peak of the curve is at a different operating point (a different optimal ratio between the input and output wind speed).
   \begin{itemize}
       \item  Along this power curve, at some operating points the CEMS exceeds the Betz power conversion  up to  $\sim$ 36\%
(Fig. \ref{fig:powerhomogeneous}). 
   \end{itemize}
   \item The CEMS also avoids the puzzling unphysical singularity in the Betz model that requires an infinitely large depleted wind field downwind as the exit velocity approaches zero.
   \end{itemize}

Instead of a thin actuator disk, picture an abstract thick harvester as in Figure \ref{fig:thickbetz}. We note that if wind cannot escape to the side of the harvester then, to conserve the (assumed uniform) flux through it's crossection, the wind speed at the outlet cannot be lower than the inlet. Indeed, most derivations of Betz law strictly require the inlet and outlet speeds to be identical.  To maintain this windspeed, Bernoulli's law  requires something to prevent the wind from expanding as it passages the actuator body -- hence our description of this as a virtual cowling assumption. (See red text in Figure \ref{fig:thickbetz})  Moreover, without a wind speed differential or crossectional change, no kinetic energy can be extracted within the harvester, and so sole source of extractable energy is a pressure differential across the harvester. Consequently, the Betz model can universally collapse any harvester down to an idealized actuator disc in which only the inlet and outlet pressure differential matters. Intuitively, this appears contrary to actual aerodynamic mechanisms in some wind harvester designs, putting the universality of Betz's law into question.

Conversely, wind expansion and wind speed variation is allowed inside the harvester in the Continuous Energy and Momentum Schema (CEMS). (see Figure \ref{fig:cemsmodel}) This is not only more intuitively comforting but gives an added source of energy extraction that allows its higher upper bound. In this schema, the Betz model becomes a special case with lower performance due to seemingly unrealistic constraints on the mechanics.  

These notional \textit{gedanken}  concepts are made mathematically rigorous and quantified in the body of the paper. We derive a new power extraction curve for the CEMS directly from Euler's law.  We confirm this bound transparently with a special case where the power factor can be computed algebraically  and its limit determined by inspection. (i.e. without the obfuscation potential of Libnetz calculus or Euler laws.) We identify the mathematical step where a Betz's law derivation implicitly imposes the limiting physical restrictions.  We also construct a continuum of harvesters that go continuously between the restricted Betz configuration to the Continuous Energy and Momentum Schema.    

Like Rankine's "actuator disc,"  the new model is a  1-D model of an idealized machine with no assumptions about how it extracts energy. Even so, it gives insight into how turbines might be improved. Like the Betz model, the CEMS is a 1-D model and does not consider, gravity, thermodynamics, angular momentum,  radial velocities, aerodynamics, blades, vorticity, or specify any mechanical mechanism. It will also ignore forces normal to streamlines just as any 1-D formulation, including the Betz model must.  Augementing Betz's law with radial or tangential flow  as well as non-uniform axial velocity has been considered previously as a consequence of including angular moment\cite{Sharpe2004} in a 2-D or 3-D model, but we show that in just 1-D, any lateral extrusion of flux out of the harvester cross-section can increase the efficiency regardless of including angular momentum and without the need for explicit multi-dimensional velocity distributions.\citep{sorensen_2015, kuik_2017}

The philosophical value of the our new universal model may ultimately exceed the worth of its higher efficiency bound-- after all, higher order  and aerodynamic effects will erode the ideal performance of any 1-D momentum and energy model.  The generalization reveals optimization principles that point in exactly the opposite directions than the actuator disc model recommends. We will briefly discuss the following insights:
\begin{itemize}
    
    \item The Betz model is optimal when it maximizes the positive and negative pressure differentials but the new model is optimized when these are minimized
    \item CEMS offers lower interference between windmills in a wind farm compared to a Betz optimal windmill
    \item Suggests use of turbines with fewer blades or lower speeds over longer axial dimension.
     \begin{itemize}
     \item Consequently, a potential for reduced turbulence, and reduced tip speed enables longer blade lengths
    \item Permits lower strength materials than required by high pressure differential Betz optimal windmills.
    \item Reduced hyperbaric embolisms bats and birds.
    \end{itemize}
\end{itemize}

Lastly, we revisit the meaning of a power extraction coefficient.  Canonically, this is ratio of the power extracted to the power in the undisturbed wind in a cross-sectional area the same size as the harvester.  However, all wind harvesters leave a "dead zone" of reduced wind velocity in a wake larger in cross-sectional area than the harvester.  When considering a compact farm of many individual harvesters, then the coefficient one might care about is ratio of the work extracted to the power \textit{within the zone of depleted wind velocity}.  Accordingly, in Appendix C, we derive optimal performance parameters for this alternative metric of areal efficiency and find the optimal power production of the actuator disc model is even lower while the CEMS is significantly higher

\begin{figure}[htbp]
\begin{center}
\includegraphics[scale=0.5]{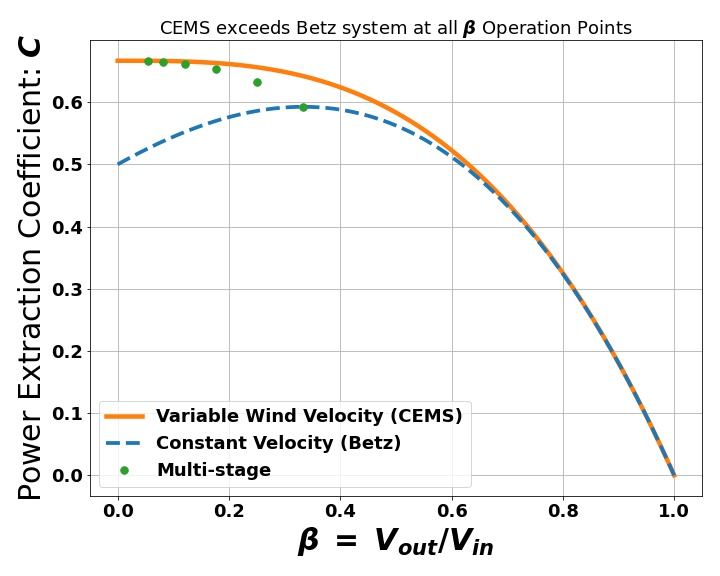}
\end{center}
\caption{\textbf{The key result of this work}: the  power extraction factor $C$ is the ratio of the power extracted to the power in an undisturbed wind ($v_{in}$) with the same cross-section as the harvester ($S$). The plots show $C$ versus   operating parameter $\beta\equiv v_{out}/v_{in}$. The CEMS (orange line) is everywhere superior to (above) the Betz curve (blue dash), indicating more power is extracted from the same area input wind-field.  The maximum of the CEMS is  2/3 of the wind power and the maximum of the Betz curve is 16/27 (59\%).   The dotted lines show how incrementally stacking 2, 4, 8, 16 or 32  Betz-stages in series ( Fig. \ref{fig:betzstages}) transforms operating point maxima above the Betz law limit and approaches the CEMS limit.  The expansion of the wind-field at $\beta=0$ is finite ($S\sqrt{3})$ for the CEMS but infinite for the Betz model.  These plot equations \eqref{eqn:ccems},\eqref{eqn:cstackbeta}, and \eqref{eqn:cbetz}.}
\label{fig:keyresult}
\end{figure}

\section{Beating the Lancaster-Betz model}
\subsection{ The Betz model}
Since the Betz model is generally well known, we defer its derivation to later in the paper.  Briefly then, the Betz actuator disc the model has a clean separation of the iso-energy expansion zone and the work extraction zone.   The constant velocity requirement across the actuator region means:
\begin{enumerate}
    \item In the Betz model, the only source of  work is the pressure drop across the actuator.
    \item The exploitable pressure change happens entirely in the iso-energetic expansion zones before and after the harvester. 
\end{enumerate}

\subsection{The Continuous Energy and Momentum Schema (CEMS)}

Unlike the Betz model, the scheme shown in Figure \ref{fig:cemsmodel} allows a continuously variable wind velocity \textit{within the harvester}.   We also extend the harvester's abstract actuator  region to cover the expansion regions of the wind, making the inlet and outlet pressure ambient, and therefore the inlet and outlet velocity are the initial and final wind speeds.  (Section 5.4  will later relax that requirement as well.)
 
In order for the velocity of an incompressible flow to vary within the confines of a fixed cross-section, the harvester must shed mass flow out of its cross-section. In Figure \ref{fig:cemsmodel}, we show this as extruded wind outside the harvester aperture where it no longer can interact with the energy extraction mechanism within. 

\subsection{Objective and outcomes}
Our goal here is to find the ratio of power extracted from a wind-field to the power in an undisturbed wind-field the same cross-section as the harvester.  One might suspect that by removing the constraints of Betz model that perhaps the answer will degenerate to having an complex dependence on the now-variable internal velocity. Or one might fear  that the new ideal harvester will have infeasible properties like infinite expansion of the wind at its best operating point, or be optimal only at an infinite length in the harvester.  In fact we shall see there is a higher but finite limit, there is no required length, and that the  downwind expansion is finite and far less than the Betz harvester requires, and, surprisingly,  it is independent of the internal velocity trajectory.  Our ultimate power curve results are summarized in Figure \ref{fig:keyresult}.

\section{Mathematical derivation of the power factor for the CEMS harvester}
\subsection{Force and Power}
The momentum of a mass flow  changes only when  force is applied, and by equal and opposite reaction we can extract work by slowing the flow.  We can thus upper bound the power extraction of any possible machine simply by the amount of power needed to slow the wind to a given value, in a way consistent with flux conservation.

In conserved incompressible flow without a force, there is no change in velocity. By Euler's theorem, the  infinitesimal velocity change from an infinitessimal force is:
\begin{equation}
dF= \hat{m}dv
\end{equation}
where the 1-D  mass flux is defined as:
\begin{equation}
   \; \hat{m} \equiv \frac{\partial m}{\partial t} = density\times (cross section)\times velocity 
\end{equation}
In incompressible flow, the density $\rho$ is simply constant scaling factor. Also due to incompressibility, the mass flux $\hat{m}$  in the intercepted wind is conserved at every plane transverse to the axis from start to finish.  Thus a  flow $\hat{m} = \rho S v_{in}$ at an inlet with cross-section $S$ with velocity $v_{in}$ subsequently requires expanding the wind cross-section inversely as the velocity changes along the harvester axis to remain constant. However, when it expands beyond the harvester's physical boundary cross-section $S$, only the portion of this mass flow resident within harvester's cross-section can receive a back-force and transfer Power (work). This interior portion of the flux $\hat{m}_{inside}$ is thus a function of the axial position-dependent velocity:
\begin{equation}
\hat{m}_{inside}(.) =\rho Sv(.)
\end{equation} 

The power needed for a change in the flux velocity is  the force times velocity.
 $$Power = Velocity\;\times \; Force$$
thus the infinitesimal power creating a infinitesimal velocity change is

\begin{equation}
\begin{split}
dP_{wind} &=v\; dF\\
    &= \hat{m}_{inside}v\;dv\\
    &= \rho S v^2\;dv
 \end{split}
\end{equation}
Where we have substituted in the velocity-dependent mass flux in the last step. Next we integrate the power expression over $\mathit{d}v$ from the inlet velocity $v_{in} $ to the exit velocity $v_{out}$ giving.  
\begin{equation}\label{eqn:pintegral}
\begin{split}
P_{cems} &= -\int_{v_{in}}^{v_{out}} dP_{wind} \\
&= -\int_{v_{in}}^{v_{out}} \rho Sv^{2}\:dv\\
&= \frac{\rho}{3}S(v_{in}^3-v_{out}^3)
\end{split}
\end{equation}

The input and output velocities here are measured at the boundary conditions where the pressure has returned to ambient so there is no pressure drop across the harvester that could supply added power. (This is not to say that the unknown mechanism of work extraction inside the device doesn't interconvert velocity and pressure as needed.) 

We substitute in a dimensionless parameter $\beta_{cems} = v_{out}/v_{in}$, which will become the design parameter we will optimize for maximum power.

\begin{equation}
P_{cems} = \frac{\rho}{3}Sv_{in}^3(1-\beta_{cems}^3)\label{eqn:pcems}
\end{equation}

 \subsection{The Conversion Factor}
 Unimpeded wind passing through a cross-section equal to $S$ carries a wind power of $\rho Sv_{in}^3/2$.  The ratio of the extracted  power to the undisturbed wind power as a conversion factor (or efficiency) is:
 \begin{equation}
\boxed{ C_{cems} = \frac{2}{3}(1-\beta_{cems}^3)}\label{eqn:ccems}
 \end{equation}
 and by inspection this is maximized at $\beta_{cems}=0$, giving our new limit on maximum conversion as 
 \begin{equation}\label{eqn:cmax}
\boxed{ C_{maximum} = 2/3}
 \end{equation}
 
\subsubsection{Momentum is balanced by the extruded wind} 
We note that at $\beta_{cems}=0$, then by definition the output velocity ($v_{out}$) is zero. However, this doesn't mean the all the wind stopped flowing. It means that all of the wind was extruded out of the harvester region $S$, leaving none to flow out behind the harvester.  The wind extruded outside this region is still flowing and carries the missing 1/3 of the original kinetic power and all of the original mass flux.  If the extruded wind happened to be traveling uniformly at ambient pressure then this would occupy an area of:
\begin{equation}S_{ext}= S\sqrt{3(1-\beta_{cems})^3/(1-\beta_{cems}^3)}= S(1-\beta_{cems})\sqrt{3/(1+\beta_{cems}+\beta_{cems}^2)}
\end{equation}
outside of the harvester with a speed of 

\begin{equation}
v_{ext} = v_{in}(1-\beta_{cems})/S_{ext}.
\end{equation}

At the optimum power conversion point $\beta_{cems}=0$,  giving  $S_{ext}= S\sqrt{3}$ and $v_{ext}=v_{in}/\sqrt{3}$. However, this model does not impose a certain spatial location or uniformity, so those external values are for intuitive convenience to picture the downwind interference this wind harvester might have.  We will compare these to the relative expansion downwind of the Betz model later. 
\subsubsection{The average wind}
\begin{figure}[htbp] 
\begin{center}
\includegraphics[scale=0.4]{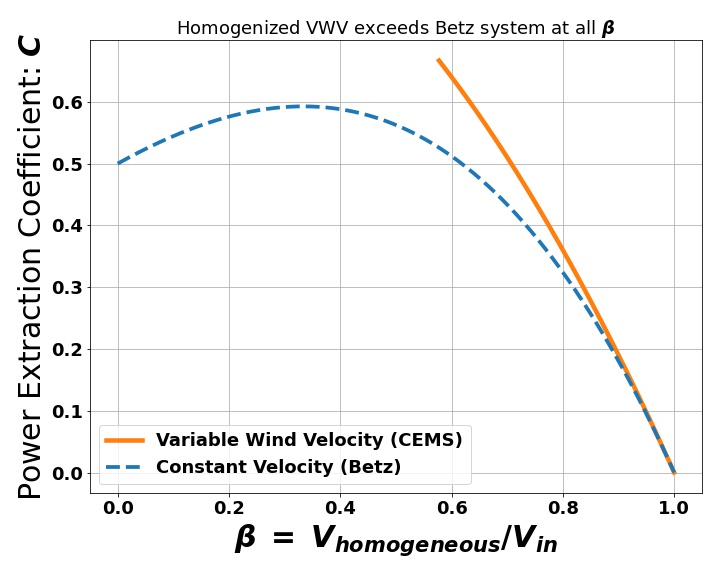}

\end{center}
\caption{The Power Extraction Coefficient is the power harvested relative to the power in the undisturbed wind-field over an area  equal to harvester's physical cross-section. This plot is the same data as figure 3 but the x-axis is now   $\beta_{homogeneous}$, the average  (transversely uniform) down-wind velocity.  \textbf{Blue Dashed} Betz model performance limit.  \textbf{Red line} VDV performance.   The CEMS curve's $\beta$ support terminates at the maximum power extraction because at that point all of the wind has been extruded from the harvester and the $\beta$ cannot go lower. As in Figure 3,  the CEMS curve is superior to the Betz curve, producing more power, and consequently the maxima are at different values of $\beta$.  }
\label{fig:powerhomogeneous}

\end{figure}

Another useful intuition comes from considering the average velocity after hypothetically re-combining the extruded wind with the wind emitted out the back of the harvester into one homogenous flow. The merged total cross-sectional area conserving power and flux is 
\begin{equation}
\begin{split}
    S_{homogeneous}& = S/\beta_{homogeneous}  \\
  v_{homogeneous} &= \beta_{homogeneous}v_{in}  \\
 \text{where}\\
 \beta_{homogeneous} &\equiv \sqrt{(1+2\beta_{cems})/3} \\
 \end{split}
 \end{equation}
 The interpretation of the artificial $\beta_{homogeneous}$ is analogous to  the homogeneous-by-definition $\beta_{betz}$. For a better comparison, the power curves in Figure \ref{fig:keyresult} are replotted in Figure \ref{fig:powerhomogeneous} using these homogeneous values on the x-axis. While $\beta_{cems}$ ranges from 0 to 1, the corresponding $\beta_{homogeneous}$ ranges from  $\sqrt{1/3}$ to 1. Thus in Figure \ref{fig:powerhomogeneous}, the orange CEMS line cannot reach below $\beta_{homogeneous}=\sqrt{1/3}$. 
 
 Since $\beta_{homogeneous}$ can never reach zero, the average downstream velocity in never zero, and so the areal spread of slowed wind downwind is finite. In contrast in the Betz model, the down wind velocity can approach zero and so to conserve mass flux the slowed wind field expands laterally to infinity. That is, in the Betz model, all the wind in the world stops!

 \section{Reality Check: a specific implementation achieves 2/3}
When tilting at a beloved 100-year-old windmill law, there's the risk of being unhorsed by an error. The differential integration  is sufficiently opaque that one might fret whether it's possible the justification of Euler's law was in some elusive way violated at the curved streamlines or control volumes.\citep{hansen_2007,sorensen_2015} Therefore we will now provide reassurance via a simpler algebraic derivation of a special case that is transparent and requires no calculus. 
\begin{figure}[htbp]
\centering
\includegraphics[scale=0.25]{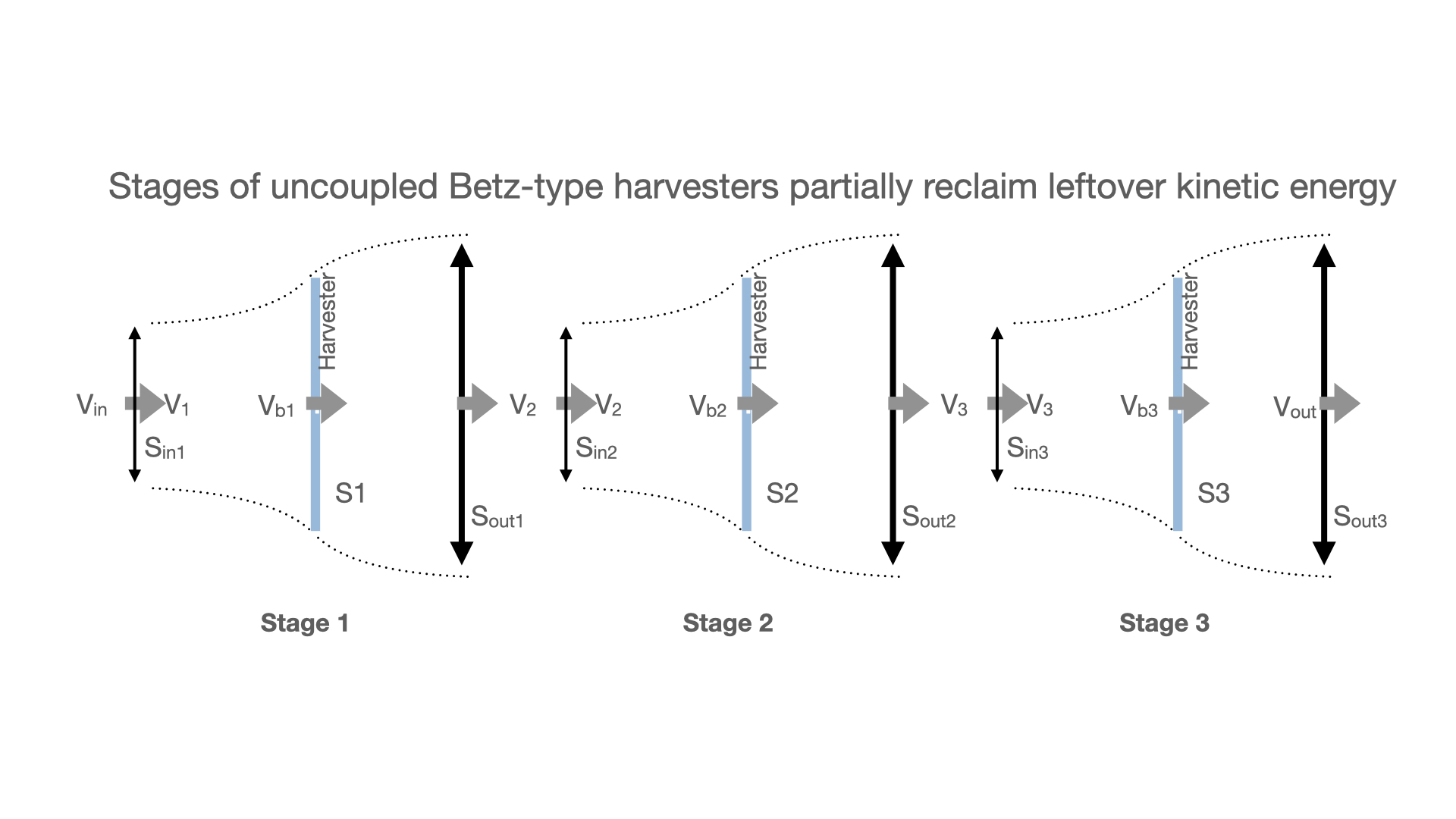}
\caption{Several Betz-like stages have been places in series.  Each stage expands and slows the wind.  Since the expansion zone is larger than the harvester diameter ($S$), it is larger than the intake region of the next stage,  and thus it is processing a portion of the original mass flow, not a new fresh part of the wind-field. To make analysis easy the stages can be placed far enough apart that between them the pressure has returned to ambient and so the streamlines are parallel again. If the first stage is operating at the Betz limit, then any additional power produced by the subsequent stages, no matter how little, is more power extraction from a wind-field that the Betz law allows. If the multiple stages are viewed as one single wind harvester then this has seemingly violated the law.  The resolution of this paradox lies in the Betz's implicit restriction to a constant velocity profile inside the harvester.  Here the velocity of the air between the stages is changing. Thus one can see that the Betz restriction on constant and uniform airflow has placed a strong constraint on what kind of physical device the Betz law applies to.   }\label{fig:betzstages}
\end{figure}

\begin{figure}[htbp]
\centering
\includegraphics[scale=0.4]{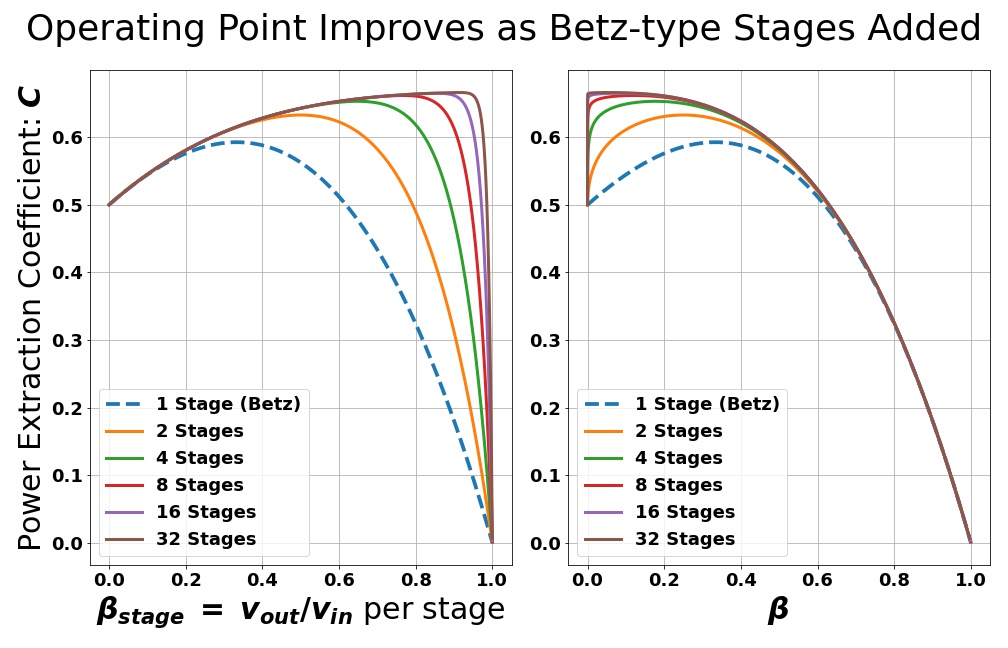}
\caption{The power factor as a function of the $\beta_{stage}$ for a series of Betz-style stages is shown.  Each curve is a different number of stage. The dashed curve is the single stage Betz device so it's the same as the Dashed curve in Figure \ref{fig:keyresult}.  As the number of stages rises, the position and value of the maximum power extraction changes.  After a few stages the maximum has nearly reached the asymptotic maximum value of 2/3. The ratio of the initial and final velocity of the system $\beta = \beta_{stage}^k$, where the maximum occurs is  approaching zero as stages are added (right plot). Conversely, the location of the maximum is at an increasingly large value of $\beta_{stage}$, which is the ratio of the input and output velocities across \textit{each} stage (Left Plot). Thus it favors decreasing wind resistance in each stage.  The dotted  points in Figure \ref{fig:keyresult} correspond to peaks of the stage models.  }\label{fig:betastage}
\end{figure}

 For this, we stack a set of Betz-type actuator stages in series, as in Figure \ref{fig:betastage}.  We  tacitly take the well-accepted  Betz power factor for each actuator disc stage as correct (see Eqn.\eqref{eqn:cbetz}  or Appendix A for derivation): 
 $$C_{betz}(\beta) = (1+\beta)^2 (1-\beta)/2$$ 
 As long as we place these stages sufficiently far apart so that the wind-field has returned to a steady velocity (and parallel streamlines) between stages, then there is no physical or mathematical coupling between these stages. With no further consideration of fluid physics,  we can simply sum up the power of serial, decoupled, independent, stages acting on the sole input wind column.  To compute the power yield for each stage,  we simply scale the input wind speed of each successive machine to the output of the prior one (i.e. multiply by $\beta_{stage}$) and apply by Betz power factor $C_{betz}$: 
\begin{equation}
\begin{split}
   P_{stack}\;\; = \;\;&\frac{{\rho}S}{2}  v_{in}^3 C_{betz}(\beta_{1})\;\;\; + \;\;\;\frac{{\rho}S}{2}(\beta_{1}v_{in})^3 C_{betz}(\beta_{2})\;\;\; + \;\;\;\frac{{\rho}S}{2}  (\beta_{1}\beta_{2}v_{in})^3 C_{betz}(\beta_{1}\beta_{2}) + \;\;\;...\\
 \end{split}  
 \end{equation}
 
 Substituting $C_{betz}$ gives:
 \begin{equation}
 \begin{split}  
    P_{stack} \;\;=\;\;& \frac{{\rho}S}{2}  v_{in}^3 (1+\beta_{1})^2 (1-\beta_{1})/2 + \;\;\;\;\;\;\;\;\;\;\;\;\;\;\;\;\;\;\;\;\;\;\; \;\;\;\;\text{Stage 1} \\
      & \frac{{\rho}S}{2} (v_{in} \beta_{1})^3  (1+\beta_{2})^2 (1-\beta_{2})/2 +\;\;\;\;\;\;\;\;\;\;\;\;\;\;\;\;\;\  \text{      Stage 2}\\
      & \frac{{\rho}S}{2} (v_{in} \beta_{1}\beta_{2})^3  (1+\beta_{3})^2 (1-\beta_{3})/2 + ... \;\;\;\;\;\;\;\;\;\;\;\;  \text{Stage 3 ...  and beyond}
\end{split}
\end{equation}
 As the equation is written it permits different ratios of input and output velocities for each stage  ($\beta_{1}, \beta_{2}, ...$), turning its optimization into a calculus of variation problem.  But to keep this algebraic,  we shall choose  the same $\beta$ for all stages.  After $k$ stages we have:
  \begin{equation}
\begin{split}
   P_{stack} \;\; &=\;\;  \frac{\rho}{2} S v_{in}^3 (1+\beta_{stage})^2 (1-\beta_{stage})( 1 +  \beta_{stage}^3  +\beta_{stage}^6 +...+ \beta_{stage}^{3(k-1)})/2\\
     &= \;\;  \frac{\rho}{2} S v_{in}^3 (1+\beta_{stage})^2 (1-\beta_{stage}) (1-\beta_{stage}^{3k})/(1-\beta_{stage}^3)/2\\
     &=\;\;   \frac{\rho}{2} S v_{in}^3 (1+\beta_{stage})^2 (1-\beta_{stage}^{3k})/(1+\beta_{stage}+\beta_{stage}^2)/2\\
\end{split}
\end{equation}
And so the power factor is:
 \begin{equation}\label{eqn:cstack}
C_{stack}(k) = \frac{P_{stack}}{\frac{1}{2} \rho S v_{in}^3} = \frac{1}{2}(1+\beta_{stage})^2 (1-\beta_{stage}^{3k})/(1+\beta_{stage}+\beta_{stage}^2) 
\end{equation} 

If we specify the desired downwind velocity output from the final stage is $v_{out}$ then
\begin{equation}
     \beta \equiv v_{out}/v_{in} = \beta_{stage}^k
\end{equation}
Substitution this into Eqn \eqref{eqn:cstack} gives:
\begin{equation}\label{eqn:cstackbeta}
    C_{stack}(k)  = \frac{1}{2}(1+\beta^{1/k})^2 (1-\beta^{3})/(1+\beta^{1/k}+\beta^{2/k})
\end{equation}
Equations \eqref{eqn:cstack}  and \eqref{eqn:cstackbeta} are plotted in Figure \ref{fig:betastage}. This shows a well behaved convergence as the number of stages, $k$, grows.  As  $k$  heads to infinity, the first and last factors converge in Eqn.\eqref{eqn:cstackbeta} to 4 and 3, respectively, leaving:
\begin{equation}\label{eqn:cinfinity}
   \boxed{ C_{\infty}  = \frac{2}{3} (1-\beta^{3})}
\end{equation}

Q.E.D. 

Using only the Betz power factor and then summing this term over the stages we have an ideal power factor for any number of stages and an asymptotic limit  identical in functional form, phenomena, and maximum value as the CEMS. 

We note that while our earlier differential derivation of the CEMS was based on power extraction soley from velocity variation inside the harvester, this concrete construction extracts all the power from a series of pressure drops, with no the velocity variation internal the actuator disc stages. The convergence of these opposite regimes  provides additional reassurance that our derivation is not imposing assumptions on the internal energy extraction mechanics as the Betz derivation did.

 \subsection{The stacked stages smoothly approach the CEMS}
 
This stack model provides insight.  By inspection, we see that the optimal $\beta_{stage}$  grows and approaches 1 as the number of stages, $k$, grows.  That means each stage is offering as little resistance as possible, limited only by consistency with the final desired $v_{out}$.  Consequently, the velocities at the first and last actuator discs approach the initial $v_{in}$ and final $v_{out}$ velocities.

\begin{mdframed}
Thus the  optimal stack grows to cover \textit{the entire wind expansion region},  asymptotically becoming the CEMS in Figure 2.  As it does, the maximum power factor approaches the CEMS limit at every value of $\beta$.
\end{mdframed}
 Note that our stack of actuator discs was a highly specific implementation which prescribes a velocity fall off between each stage at a constant ratio whereas a general machine might have a different trajectory for $v(x)$ or not even use actuator discs.  \textit{A priori}, there was  no assurance that the stacked stages would reach our abstract mathematical upper bound.  Fortunately, it did, thus securing this as a lower bound on our derived general upper bound (just in case our general derivation is somehow  faulty!).

 \subsubsection{Strawmen}
There are common strawmen that are pedagogically helpful to raise-and-dismiss at this point.   The above is a proof by construction that more energy can be extracted than Betz law allows.  To quash quibbles about whether there is a distinction between several independent turbines in a row and one single thick turbine, Figure \ref{fig:morph} shows a continuous morph of a thin turbine into a thick turbine then into multiple turbines on the same axis. As can be seen in Figure \ref{fig:betzstages}, following the Betz model, all the input air to each independent stage comes from air passing through the prior stage so it's also not picking up new kinetic energy from another source of wind energy.  A different quibble on "independence" might arise if the uncoupled rotors spin at different rates.  This can be quashed by noting that we may design each stage's rotor pitch or gearing so the axes can be locked together as one, or we could entirely avoid the rotation speed issue by the use of stages with something besides a rotor such as an no-moving-parts electrostatic system working with ionized air.

\section{Discussion}

\subsection{ Comparing the CEMS to the Betz law derivation}
 Here we provide a derivation for the Betz model in Figures \ref{fig:betzmodel1} \& \ref{fig:thickbetz} using mathematical steps closely paralleling our CEMS derivation.   We also provide an alternative and slightly more familiar derivation in Appendix \ref{appendix_betz} (and a contrast with a 2-stage harvester in Appendix \ref{appendix_ext}).  For even more detail, an excellent derivation of Betz law can be found in Reference \citep{Wind_energy_handbook} and one in terms of streamlines and control volumes can be found in Reference \citep{hansen_2007}.

There are three important deviations between the Betz and the CEMS power factor derivation.  First, since no mass is extruded in the actuator disc, the inlet and outlet mass flux are the same. Thus the Betz $\hat{m}_{inside}$ has no velocity dependence.  Second, since the Betz harvester velocity $v_{betz}$ is unchanging, the velocity multiplication converting force to power is by a constant not a variable. Since all the integrand terms are now constant they factor out of the power integral leaving only a bare $dv$.  This remaining integral portion evaluates trivially to $\Delta v = v_{in} - v_{out}$. 

The third difference of the Betz derivation involves setting this assumed-constant velocity's  value to the arithmetic mean of $v_{in}$ and  $v_{out}$.  Why it should be this particular value is not obvious by inspection but soundly follows from the prior assumptions of constant velocity and no flux extrusion.\citep{Wind_energy_handbook,hansen_2007,glauert,bianchi_wind_2007} This can be derived via momentum and energy conservation, as we do in Appendix \ref{appendix_betz}.  It can also be derived from observing that, because the velocity $v_{betz}$ is unchanged passing through the harvester,  the only source of power is from a fore-to-aft pressure drop;  in working out this pressure drop, one determines the  velocity inside a lossless Betz harvester is the arithmetic mean of the initial and final velocity.\citep{Wind_energy_handbook,hansen_2007}     Appendix  \ref{appendix_betz} also addresses a fallacy that $v_{betz}$ an "effective" or average velocity rather than a truly constant velocity.\\

In summary, following the same steps as Eqn.\eqref{eqn:pintegral}:

\begin{equation}
\begin{split}
P_{betz} &= -\int_{v_{in}}^{v_{out} }dP_{wind} = -\int_{v_{in}}^{v_{out}} v_{Betz} \times dF = -\int_{v_{in}}^{v_{out}} v_{betz}\hat{m}_{inside}\:dv\\
&= -\hat{m}_{inside}v_{betz}\int_{v_{in}}^{v_{out}} \:dv\;\; = -(\rho S v_{betz}) v_{betz}\int_{v_{in}}^{v_{out}} \:dv\\
&=\rho S v_{betz}^2 (v_{in}-v_{out})\\
&= \rho S\left(\frac{v_{in}+ v_{out}}{2}\right) ^2( v_{in}-v_{out})\\
&=\frac{1}{2}\rho Sv_{in}^3(1+ \beta)^2(1-\beta)/2\\
\end{split}
\end{equation}
\\and so the power factor is
\begin{equation}\label{eqn:cbetz}
\boxed{C_{betz} = (1+ \beta)^2(1-\beta)/2}
\end{equation}
\\
\\
Which has a maximum at $\beta_{betz} = 1/3$ giving the maximum value:
\begin{equation}
    \text{ Maximum: }C_{betz} = 16/27 \approx 59\%
\end{equation}

Contrasting the two models we note that the CEMS integral \eqref{eqn:pintegral} not only had a varying velocity in the integrand but also the harvester inlet and outlet are at the boundary condition where the pressure is ambient. In the Betz system the actuator disc is away from the ambient pressure boundaries, allowing the driving pressure-force to pillow-up from the internal wind resistance of the harvester.    Since wind-field expansion happens before the actuator disc is reached,  the power in the fraction of the wind passing outside is lost before the harvester has a chance to process it. Thus at higher loads the Betz model falls off  whereas the generlized CEMS extracts even more power. Likewise at the outlet, to foster a strong negative pressure, a large kinetic energy must pass the outlet unharvested in the Betz model.

 \subsection{Does Betz approximate a "thin" or planar harvester?}
No.   In the CEMS derivation, the velocity trajectory $v(.)$ over the path through the harvester doesn't matter.  This means that the thickness of any variable wind velocity profile could be infinitely thin, in principle.  

\begin{mdframed}
Thus the CEMS supersedes the Betz law in the limit of thin disc as well.
\end{mdframed}

The challenge to one's intuition is visualizing the lateral extrusion from an infinitely thin plane, which is why we prefer Figure \ref{fig:thickbetz} to Figure \ref{fig:betzmodel1}.  We note this limit-case challenge resembles a common pitfall in the application of Euler's closed path integral theorem:  one has to be careful that flux lines don't escape along the segments of the contour integral connecting the inlet and outlet surfaces. Taking the infinitely thin limit doesn't remove the edge flux but simply compacts it.  \textit{Explicitly forcing these side contours to have zero crossing flux implies a virtual cowling is present in the Betz model.}

\subsection{Does this "thin" limit eventually break down?}
Definitely.   As noted in the Prologue, both this model and Betz model have expansion regions, and that necessitates both radial velocities and radial velocity gradients neither of which are embodied in a 1-D model and require a 3-D model.  There may therefore be a limit on how swiftly one can expand the flow and the rate one can decelerate the axial velocity before these non-uniform and parasitic factors invalidate the implicit assumptions of the 1-D approximation.  We refer the reader to Conway's consideration of how to model the axial variation of radial velocity distributions under load.\citep{conway}    Thus, in practice there maybe a restriction on how thin one can make this actuator region and not exceed the justifications of the 1-D paradigm.

We note that at a sufficiently thin level, not only is radial uniformity not possible but one also must revisit viscosity, incompressibility, and thermal effects, that are completely outside these models.  For example, one might conjecture  that in thin "real" windmills perhaps viscosity would slow radial flow, and thus introduce some virtual cowling effect.  On the other hand, radial flows are noted in simulations and measurements of real windmills.  Thus we will simply state that all 1-D models are subject to breaking down and proper 3D simulations are required in some ranges.  But these considerations are outside the scope of this work which is to  compare 1-D models where uniformity is assumed in both.

\subsection{ Does Betz law really require a constant internal velocity?}
Yes. Occasionally it is suggested that perhaps $v_{betz}$ is merely a mean velocity.  Appendix A explains why it cannot be simply an "effective" mean flux velocity: the Betz law derivation strictly requires that harvester region flux is actually at a constant velocity equal to the inlet and outlet velocities.
\subsection{So what does the Betz law apply to?}

We think that by drawing the system as a thin actuator disc obfuscates this implicit cowling requirement, since it has no obvious port to extrude the mass flux. In Figure \ref{fig:thickbetz} we redrew the Betz system as a thick actuator element to reveal the implicit assumption introduced by the lack of wind expansion within the harvester.

\begin{mdframed}
Betz law only applies to a system with a full cowling or virtual equivalent to prevent the wind from expanding and thus satisfy the constant velocity and constant internal flux requirement.  
\end{mdframed}

 \subsection{Harvesters with arbitrary inlet and outlet velocities}
In the CEMS  the harvester starts and ends at ambient pressure, and the inlet and outlet velocities are the initial and final wind speeds.  We can adapt this to allow an inlet wind speed $v_{inlet}$ that is less than the initial wind speed $v_{in}$ via a hybrid model: place a single Betz actuator disc stage before the CEMS.  The Betz stage will handle a pressure drop and  downshift the velocity, in a way consistent with momentum conservation; after which, the CEMS operates on the reduced $v_{inlet}$ wind speed starting and ending at ambient pressure.    We can also add a single Betz stage to the outlet as well to move the outlet away from the ambient pressure and the final velocity.

Figure \ref{fig:numerical1} shows dotted lines for varied inlet and outlet velocities of this hybrid.  One can find operating points that continuously move between the Betz limit to the CEMS limit.  We do not know if this hybrid is the optimal machine but it will still form a lower bound on the upper bound of any machine with that inlet/outlet velocity. 

\begin{mdframed}
The fact this bound only equals the Betz limit at equal inlet and outlet velocities, and exceeds it all all other choices, suggests that the Betz limit is entirely due to the assumption of equal inlet and outlet velocities and not due to the thickness of the harvester or a particular choice of the  velocity profile within.
\end{mdframed}

\subsection{  Finite element analysis of the wind speed trajectory within the harvester}
\begin{figure}[h]
\centering
\includegraphics[scale=0.5]{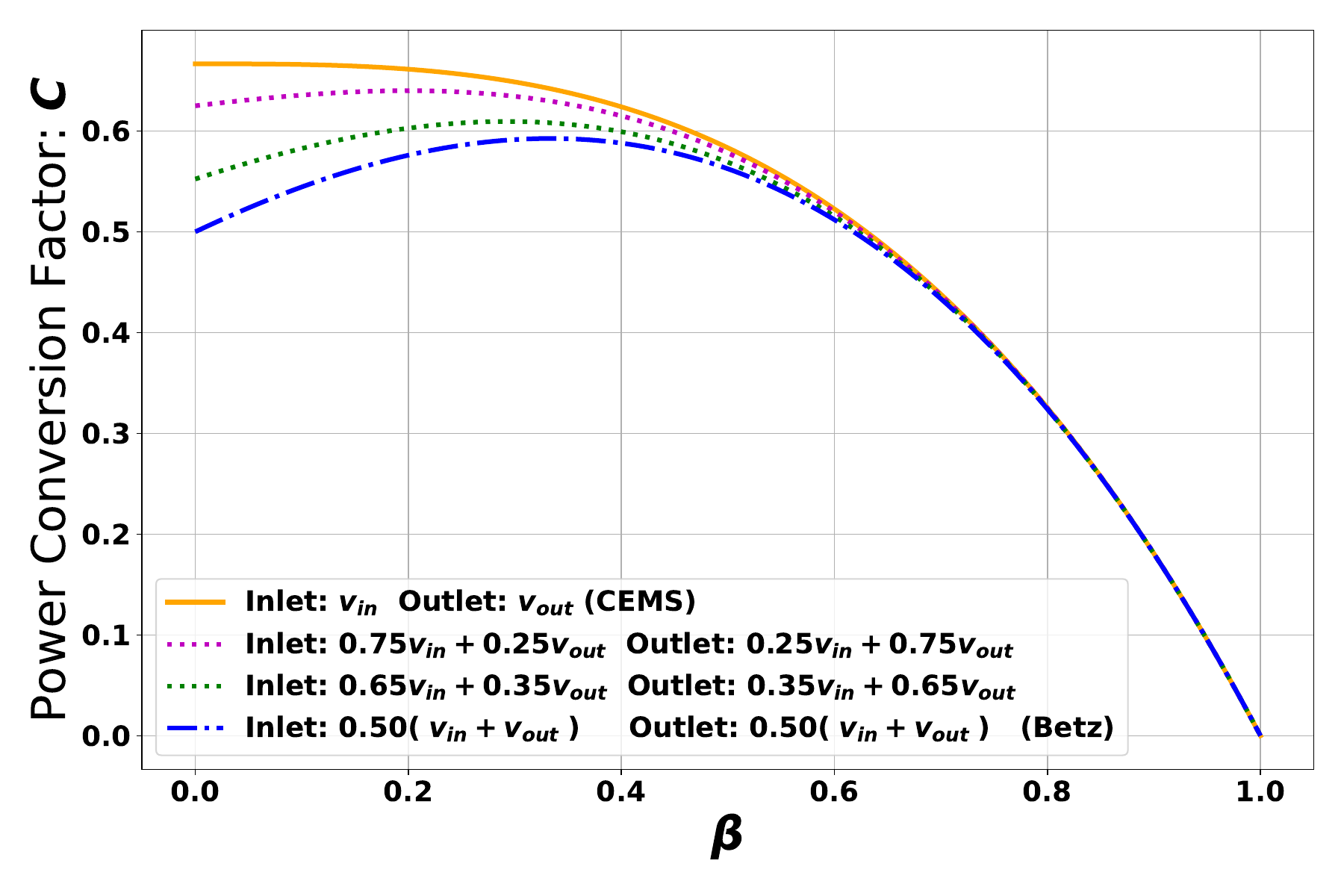}
\caption{A continuum of models vary in performance from the ideal CEMS harvester to the ideal Betz harvester.    The top-most line is the CEMS where the inlet and outlet velocities are set to $v_{in}$ and $v_{out}$.  The bottom blue line is the case where the inlet and outlet velocities are equal to the mean of  $v_{in}$ and $v_{out}$, and thus is equivalent the Betz model with no velocity variation across the harvester.  The other lines perturb the inlet and outlet velocities to values between these two limits, and get intermediate performance.  A smoothly varying velocity trajectory connects the specified inlet velocity to the specified outlet velocity, and the results are independent of this trajectory shape provided there are enough finite elements in the simulation to keep the stage to stage velocity difference small.   }\label{fig:numerical1}
\end{figure}

\begin{figure}[htbp]
\centering
\includegraphics[scale=0.5]{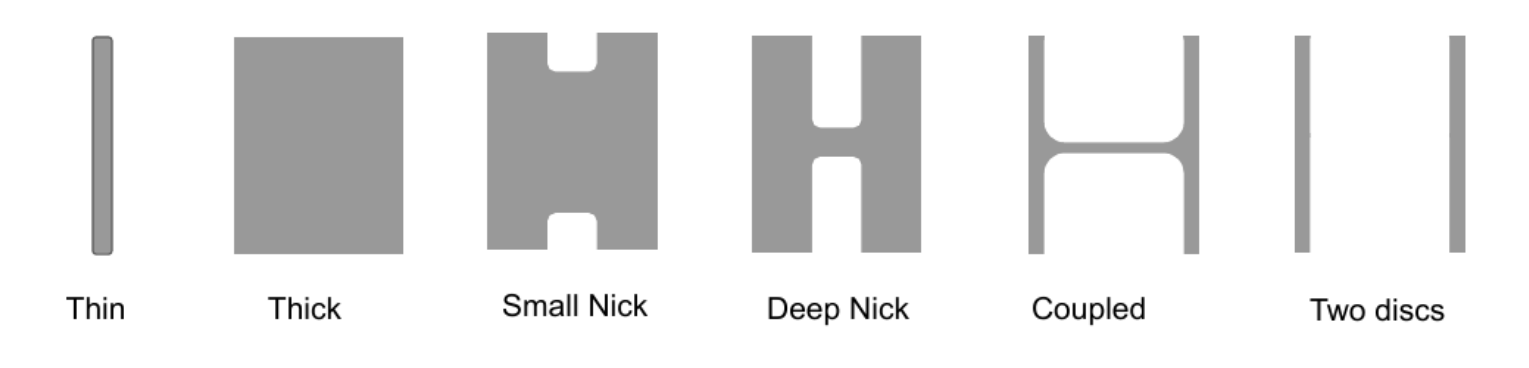}
\caption{A series smoothly morphing a single thin harvester into a thick harvester then to two thin harvesters.  The continuity of this morph shows that Betz's law, if it were truly universal, must cover thin, thick and multi-stage harvesters without exceptions.}\label{fig:morph}
\end{figure}

We are free to alter the trajectory of the velocity as it traverses the CEMS  harvester. The integral formulation provided a result independent of the velocity trajectory.  We can test this with a finite-element model.   As different models of the finite-element, we tested both  a simple extrusion differential as found in our mathematical derivation, as well as an actuator disc stage with a return to ambient pressure between each element. Both produced identical results at all operating points and agree with curves in Figure  \ref{fig:numerical1}.  Thus we confirm the velocity trajectory independence.

Caveats: As one would fully expect, the numerical simulation does require the number of finite elements to be large enough to assure a small $\Delta v$ between elements before the results converge. Amusingly, while both the math and the simulation work even if there are segments with negative expansion due to non-monotonically decreasing wind speed, that would be pulling extruded wind back in! We avoid elaborating on that here simply to avoid confusion, other than to note it is physically meaningful when instead of wind harvesting one is making a propulsion system.  And it may also be a practical approach to homogenizing  the wind speeds of the extruded and exiting wind. But those considerations are outside our scope here. 

\subsubsection {Why the Betz model is sub-optimal in performance}
An obvious questions is "if the new derivation applies to every velocity distribution, and Betz case is just a particular velocity distribution, why don't these two agree"? The intuitive answer is "if the power was shed outside the harvester or released untapped from the harvester, then the portion of the integral over that part of the velocity profile was sacrificed."

We can  verify this intuition by finite-element simulation since it book-keeps where the power loss occurs. We find an increasing amount is shed in the expansion region as the inlet velocity descends from the CEMS to the Betz case.  An increasing amount of  untapped kinetic energy exits from the outlet as we raise the outlet velocity to the Betz velocity.  

Why does the Betz model need to do that?  In the free expansion region, kinetic energy is transferred to potential energy as pressure.  The more the transfer, the higher the force on the actuator disc. But Betz can't transfer all of this to the pressure field because then the velocity through the actuator disc would be zero, making the power zero.  Therefore the Betz system compromises, and since it can't put all the kinetic energy into the pressure field it cannot access all of it.  And the higher the load the larger the defection from the CEMS upper bound.

\begin{mdframed}[]
In setting inlet and outlet velocities to the Betz  derived mean actuator disc velocity, the system must forego more energy than the minimum required by momentum conservation.\\  The CEMS sheds just the minimum required for momentum conservation and thus we believe it is the ultimate limit.
\end{mdframed}

\subsection{Why doesn't the Betz law hold for the stacked or morphed system?}
 Betz law is restricted to cases with constant velocity across the harvester actuator and without progressive mass flow extrusion during passage through the actuator.  The stacked system  violates those physical attribute restrictions.   First, between stages it bleeds mass flow out of the machine in the expansion zones, as the wind passes along the length of the stacked machine axis.  Second,  the speed of the \textit{unpressurized} wind is reducing as it passages through the harvester.  

 \subsection{Self-Consistency Check: what if we stack multiple CEMS stages?}
 Our intuition that Betz law could not hold for  all harvesters came from the observation that a second harvester placed after the first could extract power from the remaining wind. However, stacking  multiple CEMS stages wont extract more power since the optimal output wind speed is zero, leaving no more to extract.  (The extruded air conserves the mass flux even when the wind speed at the harvester outlet is zero.)
 
We also can't gain efficiency by harvesting the "extruded" wind.  For example, suppose we placed another harvester with an annular intake to capture the external airflow.  Combined, this tandem system has a total cross-section that is a factor of $1+\sqrt{3}$ larger and thus the denominator of the power factor increases.  The result is less power efficient than the first harvester alone. The same conclusion is also reached considering the smaller hypothetical homogenized cross-section because the process of capturing extruded air for such a homogenization will again increase the effective cross-section.

 \subsection{Why is the negative pressure region mandatory in the actuator disc design?}
 Although it's easy to visualize air pillowing up before the harvesters load resistance and thereby creating pressure on the actuator inlet, it's not intuitively obvious why the pressure should dip after the harvester. Indeed this was scientifically controversial from 1865 to about 1920. Ironically, in 1915 Lanchester himself didn't seem to believe his own law for that reason, and relegated it to the appendix of his article.  (If you drop it from the analysis the thrust drops by half!)\citep{Betz_1921} 
 
 So how does the Betz model impose this?  The derivation  requires that the speed of the wind exiting the actuator disc is equal to half the initial and final wind velocities. (see Appendix A) Thus it  is always higher than the final down wind velocity.  The only way free expansion of air can slow this to meet the final velocity boundary condition without violating Bernoulli's principle is for there to be a negative potential energy present in the pressure at the exit of the harvester. Reversing this logic is how  the induced pressure drop is revealed without having to know the internal operational physics of the harvester.\citep{FROUDE,Wind_energy_handbook}  Mechanically, how this happens is not specified by the model; it is just coming about because one is assuming that the final velocity might actually be achieved whether or not it actually can be. This is why it was controversial in 1915. 
 
 \begin{mdframed}
 The CEMS  does not  require  negative pressure to achieve maximum performance.
 \end{mdframed}
 
\subsection{Does the CEMS prohibit pressure variation?}
No, the stacked stage system (Figure 5) is a type of variable wind velocity harvester, and thus it is bounded by the CEMS  limit not the Betz law, even though its internal mechanism of extraction is a series of pressure drops. Thus the pressure can change within the harvester.  When running at the optimum point, the CEMS does not have any external pressure change. However, at sub-optimal operating points there can be a pressure change before or after the harvester. These were modeled in the  hybrid case shown in Figure \ref{fig:numerical1}.

\subsection{Why the CEMS is more satisfying than Betz's model}
We can now reflect on several items that make the Betz model  subjectively troubling.  To accommodate the regime where the wind velocity at the output heads to zero ($\beta_{Betz}\to0$), the cross section of the output expansion zone with nearly zero velocity must head to infinity to preserve mass flow.  That is a perplexing result: if I hold up my hand in the wind, then all wind everywhere on the earth stops blowing?   No, in reality the wind is just is extruded without expansion around the hand, just as the CEMS allows.  

In Betz model the sole means of extracting energy is via a pressure drop.  In bladed turbines there can be a pressure drop but this may not be the sole source of propulsion: Bernoulli's principle changes the air speed on opposite sides of the blade to create the lift force.  Alternatively, in a harvester without out blades such as an electrostatic decelerator using an ionized wind, one is directly slowing the ionic wind inside the harvester element.  It is far from intuitively obvious whether those mechanistic implementations can be rendered mathematically isomorphic to a simple single pressure drop; while we have not derived any specific implementation our result that more energy can be extracted than the Betz single-pressure-drop model allows is \textit{prima facie} evidence that such an isomorphism is impossible in general. Thus, contrary to claims, the Betz model implicitly restricts the physical mechanism of energy harvesting.

\subsection{Minimizing interference between nearby windmills}
Harvesters emit an expanded low velocity wind-field.  In a compact wind farm, windmills placed downwind have to be well offest transversely by more than the blade length because the foremost windmill's expansion zone will interfere not just because of turbulence but the depletion of kinetic energy.  For a given plot of land or ocean barge, this may limit how many windmills we can array without interference.  Or looking at it another way,  one large wind machine across the entire  plot of land can use all of the  wind efficiently but its expansion zone expands outside the land plot, and steals the neighboring land's access to the full wind speed.  

Here the CEMS has a design advantage.  We note that the negative pressure zone of the Betz model is external to harvester itself, and thus the expansion is not controlled and will expand uniformly in all directions.  With the CEMS, the expansion happens in the control region internal to the harvester, and so we can select the direction it is extruded.  For example, a partial cowling with an open top would extruded all the depleted air out the top of each windmill in the wind farm where it wont intersect other windmill inlets.  One still must offset the downstream windmills but only by the crossection of the windmill itself

In Appendix \ref{appendix_aerial efficiency} we consider the wake impact when such a redirection strategy is not implemented.  We compare the areal efficiency of the Betz and the CEMS.

\subsection{Reducing baric trauma in Bats and Birds}
As a matter of peripheral interest we note that autopsies of bats found under windmills show death by decompression rather than blade strikes.\citep{bats}  The region of negative pressure, at the outlet expands to far larger than the diameter of the blades, but, unlike the blades themselves, may be invisible to sight or sonar.  We note that optimizing towards Betz law maximizes this negative pressure zone, while optimizing towards the CEMS model strives for zero pressure drop at the outlet and less expansion beyond the blades.  This suggests bat-friendly windmills may not have to sacrifice performance.
 
\subsection{Prior work}
We have cited the relevant literature throughout this paper.  There are numerous discussions of augmenting  Betz law to include things like angular momentum (apropos to turbines), or to model the pressure and velocity fields in 2-D or 3-D, or to add in simple aerodynamics like rotor tip speeds or turbulence or  to bring in thermodynamics or graduate to full aerodynamics with computational fluid dynamics.\citep{Wind_energy_handbook}  All of those are outside the scope of simple 1-D models that don't make any assumptions about the mechanism. 

However, its worth pondering the turbine case where the turning blade induces angular momentum and thus partitions energy into non-axial wind velocity.\cite{kuik_2017, Joukowsky_1920} When there is no cowling, or equivalent, to provide a centriptal force, then expansion is increased by the transverse velocity. Sharpe and others have generalized 2-D and 3-D models with angular momentum for which computations show that while angular momentum saps the extractable energy, it also can slow the wind axially and thus there is an expansion from radial flow.\cite{Sharpe2004}  Under certain tip-speed conditions these contrary effects don't balance out and it might permit a (small) increase in efficiency above Betz law.\citep{Sharpe2004} This angular momentum effect has also been analyzed by Sorensen, Kuik and others.\citep{sorensen_2015, kuik_2017}.  The work in this paper shows such an effect does not require  angular momentum or vortexes or tangential flows, but is just a general consequence of mass extrusion permitting a variable wind velocity to increase efficiency inside the harvester.

Angular momentum need not be the sole mechanism for wind extrusion but its existence establishes that  wind extrusion is a common phenomena, and thus our model is not adding in some effect that doesn't naturally take place.  Measurements of "Real world" windmills show varied wind velocity just outside the radius of the blades; while the origin may be aerodynamic effects, the physical mechanism of extrusion doesn't matter to the model.

There are varied publications claiming to disprove Betz law that contain apparent errors,  (and oh how we hope we are not joining that group.)  Among these is a derivation for the actuator disc model that erroneously explored varied ways to integrate over parts of the iso-energetic Bernoulli wind expansion outside the harvester rather than over the actuator disc where the power is actually harvested. Using an indefinite integral rather than the actual boundary conditions resulted in several possible power factors lacking dependence on the output velocity, including 100\% and 67\% efficiency.  None-the-less, reverse engineering these logic mistakes proved instructive when debugging our own numerical simulations.\citep{"sen_2013"}

 \section{Conclusion}
 We have derived an upper bound for the fraction of wind kinetic energy any wind harvester can achieve.  This Continuous Energy and Momentum Schema (CEMS) relaxes the constant internal velocity restriction of the Betz law derivation.  By implication, Betz law, which is just a special case of the CEMS model, is not actually a universal law.  Since most windmills do not have cowlings and do extrude air, Betz law isn't a limit for most windmills.   Relaxing these assumptions, we can allow mass to extrude laterally out of the harvester's cross-sectional area.  Radial flow has been considered previously as a consequence of including angular momentum,\cite{Sharpe2004} but here we showed that allowing flow out of the harvester cross-section increases the efficiency without any need to consider angular momentum, radial velocity, transverse non-uniformity within the harvester cross-section, nor an explicit 2-D or 3-D model.  The new upper bound allows a theoretical 2/3rds of the kinetic energy to be extracted from a wind-field of a given cross-section and still conserve mass flow.  The point of operation for optimal performance is also different.
 
 Stacking identical Betz harvesters in series produces more power extraction than Betz Law allows and can reach the new limit.  Unlike the Betz derivation, the new derivation is self consistent, as stacking CEMS harvesters in series does not harvest a greater fraction of the wind power.   We did not explore relaxing the constraint of transverse wind-field uniformity within the harvester, as this was a 1-D model.   Unlike the Betz model, the CEMS avoids the pathological case of zero wind velocity with infinite expansion down-wind.
 
 The quotidian implication of raising the fundamental limit on power extraction from 16/27 to 2/3 is that there is more room for improvement in real-world windmill efficiency that was previously recognized. Moreover,  a design insight is that it is beneficial to minimize pressure build up.  Conversely, Betz harvesters strive for high inlet  pressure.   Because this pressure is developed outside the device, it sheds flux containing untapped energy,  and thus under-performs the CEMS design, especially at high load.  While it may be technically challenging to build a real-world harvester that eliminates all inlet pressure build up, it is useful as a new rule of thumb for design.  Avoiding a large pressure build up offers the opportunity to profit from an extended axial length harvester that can gradually develop the force and thus avoid other frequent issues in turbine windmills  such as torsional and bending forces on the blades from high pressure gradients.
 
 Furthermore since the CEMS harvester  is valid for an infinitely thin harvester as well, it supersedes Betz law in that regime as well.  Thus one should not think of Betz law as the limit case for a narrow blade windmill.  Instead Betz is a limit on wind machines specially constructed so as to not allow air to escape during passage: for example a windmill with a cowling or tunnel.   The maximum 2/3 conversion of the CEMS harvester  is correct for all  HAWT windmills within its assumption of uniform transverse internal wind speeds, regardless of the thickness of the windmill.

\section{ Appendices} 
 
\subsection{  Appendix A: Alternate derivation of Betz law}\label{appendix_betz}
Here we give a terse outline of another way of deriving Betz law that is the most common.\citep{Wind_energy_handbook}  We gave a slightly different version in the main text because it is more easily compared to our CEMS.  The problem with the following is that it obfuscates the assumption of constant internal velocity inherent in Betz law.  In Appendix 2 we will follow the same derivation as in this Appendix but explicitly break the constant velocity assumption.
\begin{enumerate}

\item Compute the rate of work done to change the flowing momentum: 
\begin{equation}
\begin{split}
    P_{work} &=  \text{ (Rate of momentum change)} \times \text{Velocity}\\
           &= \hat{m}(v_{in}-v_{out}) v_{b}
\end{split}
\end{equation}

\begin{itemize}
   \item[] Where $\hat{m}= \rho S v_b$ and $v_b$ is the as-yet unknown velocity through the actuator disc in Figure \ref{fig:betzmodel1}.
\end{itemize}
\item Compute loss of Wind kinetic energy at ambient pressure endpoints:
\begin{equation}
      P_{wind} = \frac{\hat{m} v_{in}^2}{2} -  \frac{\hat{m} v_{out}^2}{2}
\end{equation}

\item Equate theses (to assure both momentum and energy conservation) and solve for any unknowns (i.e. $v_{b}$). 
$$P_{work}=P_{wind}$$
gives the classic Froude\citep{FROUDE} result:
\begin{equation}
    v_b = \frac{v_{in}+v_{out}}{2}
\end{equation}
\item Now all the variables are know and (1) and (2) are now equal so power function is known.
\begin{equation}
P_{wind}=P_{work} = \frac{\rho S }{2}v_{in}^3 (1+\beta)^2( 1-\beta)/2
\end{equation}
\begin{itemize}
\item[] where $\beta = v_{out}/v_{in}$
\end{itemize}
which recovers well accepted equation for Betz law and identical to \eqref{eqn:cbetz}.
\end{enumerate}

\subsubsection{The fallacy of "effective" velocity}\label{appendix_fallacy}
Some derivations try to evade the necessity that Betz law assumes a constant velocity, by  claiming that $v_{b}$ is an abstract "effective" velocity representing the average flow and thus the average generation of power.  While that would be a clever way to pretend the above is valid even in the face of variable velocity, it actually invalidates the above derivation.  To see this, briefly consider an analogy of an airplane flying from point A to point B. The plane might change its speed over time, but we could compute an effective average velocity if knew how long the flight takes and the total distance; this average  would also equal the arithmetic mean of all the intermediate variable velocities over time.  On the other hand, if the amount of fuel the plane burned depended on a drag proportional to $v^2$ then we could not use the arithmetic average velocity to compute the fuel use. But if we knew the Root Mean Square (RMS) velocity, we could use that, instead of integrating, the fuel usage over the velocity profile.   

Turning to the problem at hand,  we see the kinetic energy expression for the power introduces $v_b$ linearly while the momentum expression for the power introduces it as $v_b^2$. If we want to avoid integrating both expressions over a variable velocity, could we use an "effective" velocity?  The momentum expression will require $v_b$ to be the Root Mean Square flux, and the kinetic energy expression will need $v_b$ to be the arithmetic mean  flux. Since these are not the same, dividing out the "effective" mass flux is not allowed after equating these power formulae, invalidating the above derivation.   In general, the RMS only equals the arithmetic mean when all the values are constant.  Thus resorting to an "effective" velocity is a fallacy and cannot be used to evade the restriction of this derivation to a constant velocity case. 

 \subsection{Appendix B: Derivation without assuming inlet and outlet velocities are equal}\label{appendix_ext}
 This derivation will be for a special case, specifically 2 stages of actuator discs in Fig.\ref{fig:betzstages} , and is \textit{not} intended to be a universal result.  The point here is to arrive at a power law that is different to and exceeds the Betz law limit, by using exactly the same formalism of momentum and energy conservation as Appendix \ref{appendix_betz}.  That will prove that Betz law is not universal and shows how extruded wind is missing from the Appendix \ref{appendix_betz} derivation.

\begin{enumerate}

\item Compute the rate of work done to change the flowing momentum at all points of change (specific to case of 2 stages): 
\begin{equation}
\begin{split}
    P_{work} &=  \sum{\text{ (Rate of momentum change)} \times \text{Velocity}}\\
           &= \hat{m_1}(v_{in}-v_{ext}) v_{b1} + \hat{m_2}(v_{ext}-v_{out}) v_{b2}
\end{split}
\end{equation}

\begin{itemize}
   \item[] Where $\hat{m_1}= \rho S v_{b1}$ and $\hat{m_2}= \rho S v_{b2}$.
\end{itemize}
\item  Compute loss of Wind kinetic energy at ambient pressure endpoints:
\begin{equation}
      P_{wind} = \frac{\hat{m_1} v_{in}^2}{2} -  \frac{\hat{m_2} v_{out}^2}{2} - \frac{(\hat{m}_1-\hat{m}_2) v_{ext}^2}{2}
\end{equation}

\item  Equate theses (to assure simultaneous momentum and energy conservation), and solve for the unknowns  ($v_{b1}$, $v_{b2}$, $v_{ext}$), which are respectively the first stage velocity, the second stage velocity, and the extruded wind.

However, since this is now an under-determined equation, instead of a unique solution we get a family of relationships between the unknowns.  Two family groups are easily found by inspection. \\ 
\begin{itemize}
    \item Group 1: Betz law (trivial case)\\
    $$v_{b1} = v_{b2} = \frac{v_{in}+v_{out}}{2},  \;\;\;\;\;v_{ext} = 0$$
    \item Group 2: Power greater than or equal to Betz law\\
    $$v_{b1} = \frac{v_{ext}+v_{in}}{2}, \;\;\;\;\;v_{b2}=\frac{v_{ext}+v_{out}}{2}, \;\;\;\;\; v_{in}\ge v_{ext} \ge v_{out}$$
\end{itemize}

\item The group one case has no mass extruded ($V_{ext}=0$) and so the velocity at the inlet and outlet must be equal, naturally recovering Betz law for this condition.  The Group 2 family has a maximum power output when $v_{ext}= (v_{in}+v_{out})/2$.   This means the two stages have different effective $\beta$,  whereas in that earlier k-stage derivation we choose, for convenience, to make all the stages have the same $\beta$. If we wanted the stage $\beta$ to be the same here, then we select the geometric mean $v_{ext}= \sqrt{v_{in}v_{out}}$: this case is plotted (2 stages) in Figure \ref{fig:betastage} and can be seen that it exceeds Betz law at all operating points.   The performance  of Group 2 in pessimized only at the edge cases of $v_{ext}=v_{in}$ or $v_{ext}=v_{out}$;  this worst case is simply once again equivalent to Betz law where one stage is doing all the work and the other lets the air pass.

\end{enumerate}
Thus once again, even using this alternative formalism, the Betz law underperforms.  These equations are specific to just 2 stages: we did not intend to achieve the optimal result derived in the paper in this Appendix because the algebra becomes prohibitively dense using this formalism, but even a special case is sufficient to show the assumption of equal inlet and outlet velocities can be relaxed and thus produce more power than Betz law allows.

 \subsection{Appendix C: An alternative performance metric: Aerial Efficiency}\label{appendix_aerial efficiency}
For a wind farm we need to compare the harvested power to the unavailable power given up by the inaccessible expanded wake.   An alternative efficiency metric is the ratio of extracted  power to the undisturbed wind power \textit{in a cross section the size of the fully expanded wake }, as opposed to just the smaller inlet cross section. In the case of the Betz harvester it was noted above that the expansion occurs outside the control region of the harvester.  In the case of the CEMS we have the option to re-direct the extruded air expansion above the plane other windmills lie in, minimizing the areal interference.  In that case the Aerial Efficiency is just $C_{cems}$ itself.  However, such a partial cowling might be cumbersome to implement.  Therefore in this appendix we look at the aerial efficiency in cases where that control on the CEMS wake is not implemented.

 \subsubsection{Non-homogenized CEMS wake compared to Betz wake}
  This Areal Efficiency is shown in Y-axis of both plots in Figure \ref{fig:areaeff}.  In the left hand plot the x-axis is the power captured.  Comparing the Red and dashed Blue lines one sees that in this measure, the Betz machine is superior for the same output power, but ultimately the CEMS can extend to higher output power.  The CEMS is at a disadvantage here because its output is composed of different velocities and thus it will occupy more area than a uniform flow at the same  kinetic energy and flux (uniform velocities are always more compact conduits).  
 
 \subsubsection{Homogenized Aerial Efficiency}
\begin{figure}[htbp]
\begin{center}
\includegraphics[scale=0.5]{"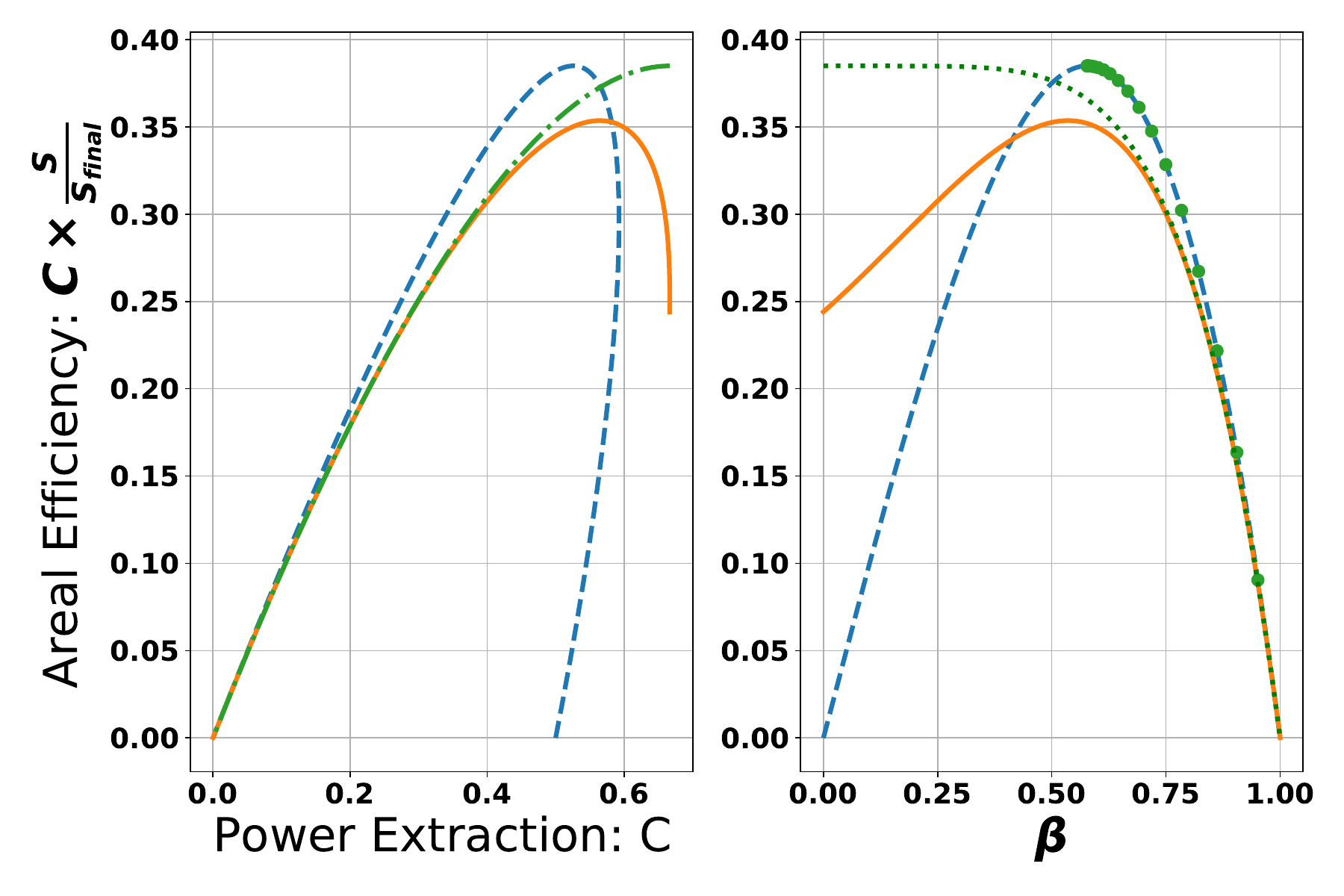"}

\end{center}
\caption{The Areal Efficiency is the power harvested relative to the power in the undisturbed wind-field over an area  equal to harvesters wake cross-section (rather than relative to the harvester's physical cross-section). \textbf{Blue Dashed} Betz model performance limit.  \textbf{Red line} CEMS performance. \textbf{Left:} X-axis is the power extraction coefficient.  The CEMS achieves a higher total power extraction but compared just over the power extraction range of the Betz machine, the CEMS  leaves a larger cross-section of disturbed air in its wake. \textbf{Right:} The same data shown versus $beta$. \\
In both figures the \textbf{Green} curves are the CEMS performance recomputed after compacting the wake by making the wind velocity transversely uniform. This matches the peak areal efficiency of the Betz curve, but can extract more power.  In the right plot the $\beta$ for the green line is displayed two ways: for green small dots the $beta$ is for the air velocity exiting the harvester element only (and thus can be compared to the red line easily). For large dots, $beta$ is computed for the final homogenized uniform velocity.  The large green dots follow the Betz Areal Efficiency curve up to their coincident maximum power point, but since this occurs when all the CEMS harvester air has been extruded, the $\beta$ cannot go lower for the homogenized flow. The respective power extraction factor limits (C) when operating at the peak aerial efficiency are 0.526, 0.564, and 0.667 for the Betz, CEMS, and Homogenized curves. }\label{fig:areaeff}

\end{figure}

However, we can improve on that.  If we concoct a mixing device to homogenize the outlet and extruded winds into a uniform velocity then we can compact the area of the wake cross section.  Hypothetically, we would obtain the green curves in Figure \ref{fig:areaeff} whose areal efficiency is better than the uncompacted CEMS curves. The green curve is plotted two ways in the righthand figure: the small dots use the original $v_{out}$ of the harvester element to compute $\beta$ on the x-axis, and the large green dots use the final velocity $v_{homogeneous}$ after homogenization with the external wind.

 In the left hand plot, the green lines show the homogeneous CEMS would achieve a maximum Areal Efficiency tied with the Betz machine, but producing more total power.   The respective power conversion limits when operating at the peak aerial efficiency are 0.526, 0.564, and 0.667 for the Betz, CEMS, and Homgenized curves in the left plot.  Thus to achieve maximal aerial efficiency the Betz model and CEMS should be operated below their peak power factor points.  However, if the airflow can be homogenized then the full peak power (2/3) of the CEMS limit is possible at maximum aerial efficiency.

  \bibliographystyle{plain}
\bibliography{betz}

\end{document}